\documentstyle[12pt,aps,prd]{revtex}
\begin{document}
\tightenlines
\title{Supersymmetric quantum solution for FRW cosmological model with
matter.}
\author{J. Socorro{\footnote{E-mail: socorro@ifug3.ugto.mx}} \\
Instituto de F\'{\i}sica de la Universidad de Guanajuato,\\
Apartado Postal E-143, C.P. 37150, Le\'on, Guanajuato, Mexico.
}
\date{\today}
\maketitle

\begin{abstract}
Using technique of supersymmetric quantum mechanics we present new
cosmological quantum solution, in the regime for FRW cosmological model using 
a barotropic perfect fluid as matter field.
\end{abstract}

\vspace{0.5cm}
\noindent {PACS numbers: 04.20.Jb; 04.40.Nr; 12.60.Jv; 98.80.Hw}

\section{Introduction}
We are interested to study some cosmological models in the 
supersymmetric quantum mechanics scheme.
Recently, particular exact solutions to the Wheeler-DeWitt (WDW) 
equation  in Witten's\cite{witten} supersymmetric quantum mechanics for all 
Bianchi Cosmological Class A Models in the  Einstein theory were 
found\cite{socorro2}. 

One goal in this work is to try to solve an ambiguity in the factor ordering 
of the position and momenta operators and give 
{\it selection rules} that fix the parameter that measure this ambiguity.
Such  ambiguities always arise, when there are expressions containing the  
 product of non-commuting quantities that depend on $q^\mu$ and $P_\mu$ 
as in our case. It is then necessary to find some criteria to know which 
factor ordering should be selected. The factor ordering in the semiclassical
approximation is irrelevant, but not so in the exact theory. In a previous
 work \cite{socorro2}  the global factor  was dropped by hand and the factor 
ordering ambiguity was avoided when they factorized the WDW equation.

Thus, the idea of Witten  \cite{witten} is to find the supersymmetric 
super-charges operators $Q$, $\bar Q$ that produce a super-hamiltonian 
$H_{ss}$, and that satisfies the closed superalgebra
 
\begin{equation}
\left\{Q,\bar Q\right\} = H_{ss}, \qquad 
\left[H_{ss},Q \right]=0, \qquad \left[H_{ss}, \bar Q \right]=0,
\end{equation}
where the super-hamiltonian $H_{ss}$ has the following form
\begin{equation}
H_{ss}:=  H_0+ 
\frac{\partial^2 \Sigma(x,y)}{\partial q^\nu \partial q^\mu}\left[
\bar \psi^\nu,\psi^\mu\right], 
\label{super}
\end{equation}
here $H_0$ is the bosonic Hamiltoniano and $\Sigma$ is known as the
super-potential term that is related with the potential term that
appears in the bosonic hamiltonian. This idea was applied 
in the reference \cite{socorro2} for all Bianchi type cosmological models.

In this approach, the hamiltonian $\rm H_{ss}=\frac{1}{2} Q^2$ is  positive 
semi-definite and a supersymmetric state with $\rm Q |\Psi>=0$ is automatically
a zero energy ground state. This simplify the problem of finding supersymmetric
ground state because the energy is known a priori and also because the
factorization of $\rm H_{ss}|\Psi>=0$ into $\rm Q |\Psi>=0$, 
$\rm \bar Q|\Psi>=0$ often provides a simple first-order equation for the
ground state wave function.
The simplicity of this factorization is related to the solubility of certain
bosonic hamiltonians. For example, in this work we find 
 for the empty (+) and filled (-) sector of the fermion 
Fock space  zero energy solution 
\begin{equation}
\rm |\Psi_\pm>= e^{\pm \Sigma}|\pm>,
\label{semi}
\end{equation}
 where $\Sigma$ denotes a superpotential, and $\Psi_+$ and 
$\Psi_-$ are the corresponding components for the empty and filled
sector in the wave function.
We also observe a tendency for supersymmetric vacua to remain close to their
semi-classical limits , because in this work and others \cite{socorro2}, 
the exact solutions (\ref{semi}) are also the lowest-order WKB aproximations.

The paper is organized in the following way. In section II, the 
ADM lagrangian of our model is constructed. In Section III, we derive the 
corresponding 
Hamiltonian that  allow us to obtain the quantum WDW equation for the 
FRW cosmological model with matter field. In section IV we derive the WDW 
solution in the supersymmetric quantum mechanics approach. Section V is 
devoted to conclusions.  

\section{ADM Lagrangian formulation}

We consider the total lagrangian, where a part is  geometry and the other 
part corresponds to the matter field. We will consider a perfect fluid with 
barotropic equation of state as our matter field.
\begin{equation}
{\cal L}_{total}= {\cal L}_{geom} + {\cal L}_{matter},
\label{lagrangian}
\end{equation}
where the lagrangian density for geometry is the usual
\begin{equation}
{\cal L}_{geom}=\sqrt{^(4)g} \, R,
\label{lagrangiangeometry}
\end{equation}
where $R$ is the Ricci scalar. 

Using the metric for FRW
\begin{equation}
\rm ds^2 =g_{\mu\nu}(x^\alpha)\, dx^\mu \, dx^\nu= 
- N^2\,dt^2 +A^2 \left[\frac{dr^2}{1-\kappa r^2}+r^2\left(d\theta^2 +
sin^2\theta d\phi^2\right)\right],
\label{metrica}
\end{equation}
where $\rm N$ is the lapse function, $\rm A$ is the scale factor of the model,
and $\kappa$ is the curvature index of the universe
( $\rm \kappa=0,+1,-1$ plane, close and open, respectively)

The covariant component for the tensor metric are

\begin{equation}
\rm g_{tt} = -N(t)^2,\qquad g_{rr}=\frac{A(t)^2}{1-\kappa r^2} ,
\qquad g_{\theta\theta} =A^2r^2,\qquad g_{\phi\phi} =A^2r^2\sin^2\theta,
\end{equation}
and the contravariant components
\begin{equation}
\rm g^{tt} =- \frac{1}{N^2} , \qquad g^{rr}= \frac{1-\kappa r^2}{A^2} , 
\qquad g^{\theta\theta} =\frac{1}{A^2r^2}, 
\qquad g^{\phi\phi}=\frac{1}{A^2r^2sin^2\theta}.
\end{equation}

With these elements, we can calculate the nonzero Chistoffel symbols
\begin{eqnarray}
{\rm \Gamma^t_{tt}}&=&{\rm \frac{\dot N}{N}  ,\qquad 
\Gamma^r_{tr}=\Gamma^\theta_{t\theta} = \Gamma^\theta_{\theta t}= 
\Gamma^\phi_{t\phi}=\Gamma^\phi_{\phi t} =\frac{\dot A}{A} } 
\nonumber \\
{\rm \Gamma^t_{rr}}&=&{\rm \frac{-A\dot A}{N^2\left(\kappa r^2-1\right)},
\qquad \Gamma^r_{rr}= \frac{-\kappa r}{\kappa r^2-1} ,\qquad 
\Gamma^t_{\theta\theta}= r\left(\kappa r^2-1\right) } 
\nonumber \\
{\rm \Gamma^\theta_{r\theta}}& =&{\rm \Gamma^\theta_{\theta r}=
\Gamma^\phi_{r\phi} =\Gamma^\phi_{\phi r} =\frac{1}{r}, 
\qquad \Gamma^t_{\theta\theta} = \frac{Ar\dot A}{N^2}, 
\qquad \Gamma^\theta_{\phi\phi}= -cos\theta sin\theta } 
\nonumber \\
{\rm \Gamma^\phi_{\phi\theta}}& =&{\rm \Gamma^\phi_{\theta\phi} =
\frac{cos\theta}{sin\theta},\qquad \Gamma^t_{\phi\phi}=
\frac{Ar^2\dot A \sin^2\theta}{N^2},\qquad \Gamma^r_{\phi\phi}=
r\sin^2\theta \left(\kappa r^2-1\right)},
\label{chris}
\end{eqnarray}
where $\rm \dot A =\frac{d A}{d t}$.

The Ricci scalar becomes

\begin{equation}
\rm R=- \frac{6}{AN^2}\frac{d^2 A}{dt^2} -\frac{6}{A^2 N^2}
\left(\frac{dA}{dt}\right)^2
+ \frac{6}{AN^3} \frac{dA}{dt} \frac{dN}{dt} - \frac{6 \kappa}{A^2}.
\label{escalar}
\end{equation}

We consider a perfect fluid energy-momentum tensor
\begin{equation}
\rm T_{\mu\nu} = pg_{\mu\nu} +\left(p+\rho\right)U_\mu U_\nu,  
\end{equation} 
where $\rm p$, $\rho$, $\rm U_\mu$ are the presion, energy density and the
four-velocity of the system, respectively.
Using the covariance of this tensor  
\begin{equation}
\rm T^{\mu\nu}_{\quad;\nu} = 0,
\end{equation}
we obtain the following partial differential equation 

\begin{equation}
\rm 3\frac{d A}{d t}\rho + 3\frac{d A}{d t}p +\frac{d \rho}{d t}A = 0,
\end{equation}
and using the barotropic state function between the presion and the energy
density, $ p =\gamma \rho$, with $\gamma$ a constant, we have the solution 
for the energy density as a function of the scale factor of the FRW universe
\begin{equation}
\rm \rho = \frac{M_\gamma}{A^{3 \left(\gamma+1\right)}},
\end{equation}
where $M_\gamma$ is a integration constant.

Using the line element for FRW, the density lagrangian for geometry
has the following structure
\begin{eqnarray}
{\cal L}_{geom} &=& {\rm \sqrt{-^{(4)}g} \, R = 
- \frac{6A^2}{N}\frac{d^2 A}{dt^2} -\frac{6A}{N}
\left(\frac{dA}{dt}\right)^2
+ \frac{6A^2}{N^2} \frac{dA}{dt} \frac{dN}{dt} - 6 \kappa N A }
\nonumber\\
&=& {\rm \frac{d}{dt} \left( \frac{-6A^2 \dot A}{N} \right) + \frac{6A}{N}
\left(\frac{dA}{dt}\right)^2 -  6 \kappa N A}
\end{eqnarray}
and the matter density lagrangian\cite{Ryan,pazos}
\begin{eqnarray}
\rm {\cal L}_{matter}&=&-16 \pi N  \rho \left\{ (\gamma +1) 
\left(1+g^{km} \, U_k\, U_m\right)^{\frac{1}{2} } 
- \gamma  \left(1+g^{km} \, U_k\, U_m\right)^{-\frac{1}{2} } \right\} 
\nonumber\\
&+& 16\pi \rho (\gamma +1) U_m N^m.
\end{eqnarray}
In ${\cal L}_{matter}$ we consider the comovil fluid  ($U_k=0$), and the gauge 
$\rm N^k=0$, obtaining
\begin{equation}
{\cal L}_{matter} = 16\pi NM_\gamma A^{-3 \gamma}.
\end{equation}
Finally, the total density  lagrangian has the following form
\begin{equation}
{\cal L}_{tot} =\frac{d}{dt} \left( \frac{-6A^2 \dot A}{N} \right) +
 \frac{6A}{N}\left(\frac{dA}{dt}\right)^2 -  6 \kappa N A + 
16\pi G N M_\gamma A^{-3 \gamma}.
\end{equation}
\section{Hamiltonian formulation}
Following the well-known procedure for obtaining the canonical hamiltonian 
function, we define the canonical momentum conjugate to the generalized 
coordinate A (scale factor) as 
$\Pi_A \equiv \frac{\partial L}{\partial \dot A}$

\begin{equation}
 L =6\left[\frac{A}{N}\left(\frac{dA}{dt}\right)^2 -  N \kappa  A + 
16\pi G N M_\gamma A^{-3 \gamma}\right],
\end{equation}
 or in its canonical form
\begin{equation}
L = \Pi_A\dot A - N H=\Pi_A\dot A - N\left[\frac{\Pi ^2_A}{24A} +6\kappa A -
16\pi G M_\gamma A^{-3\gamma}\right],
\label{lagrangiano}
\end{equation}
where
\begin{equation}
\rm H =\frac{\Pi ^2_A}{24A} +6\kappa A - 16 \pi G M_\gamma A^{-3\gamma},
\label{hamiltoniana}
\end{equation}
when we perform the variation of this lagrangian (\ref{lagrangiano}) 
with respect to N,  $\frac{\partial L}{\partial N}=0 $, implying $ H =0$. 

The quantization procedure will be made in the usual way, considering the
momentum as operators and taking the following representation for them,
but it is possible to realize other type of quantization for this same
model, for example, the supersymmetric quantum mechanics scheme
\cite{rs1,rs2,rosu}.
\begin{equation}
 H \rightarrow \hat H\Psi= 0,   \qquad  
\Pi_A \equiv -i \hbar \frac{\partial}{\partial A} ,
\end{equation}
where $\Psi(A)$ is the wave funtion of the FRW universe model. In all the work
we take $\hbar=1$.

With this assumptions, (\ref{hamiltoniana}) is transformed in a non lineal 
differential equation.
\begin{equation}
\rm \hat H = \frac{1}{24A}\left[-\frac{d^2}{dA^2} +144\kappa A^2 -
384\pi G  M_\gamma A^{-3\gamma +1}\right].
\label{hamiltonian1}
\end{equation}

In \cite{rs2} was shown that closed, radiation-filled FRW quantum universe
for arbitrary factor ordering obey the Whittaker equation.

One important results yields at the level of WKB method, where we  do 
the transformation 
$\Pi_A\rightarrow \frac{d \Phi}{d A}$, then (\ref{hamiltoniana}) is 
transformed in the Einstein-Hamilton-Jacobi
equation, where $\Phi$ is the superpotential function, that is related 
to the physical potential under consideration.  

Introducing this ansatz in (\ref{hamiltoniana}), 

\begin{equation}
\rm H = \frac{1}{24A}\left[\left(\frac{d\Phi}{dA}\right)^2 +144\kappa A^2 -
384\pi G  M_\gamma A^{-3\gamma +1}\right],
\end{equation}
thus, the superpotential $\Phi$ have the following form
\begin{equation}
\Phi = \pm\int{\sqrt{384 \pi G  M_\gamma A^{-3\gamma+1}-144\kappa A^2 }}\,dA,
\label{phi}
\end{equation}
where for whatever $\gamma$ the integral has the solution
\begin{eqnarray}
&&\int \sqrt{384 \pi G M_\gamma A^{-3\gamma + 1} - 144 \kappa A^2} dA= 
\sqrt{-144 \kappa A^2  + 384 \pi G  M_\gamma A^{1 - 3\gamma} } \left\{
\frac{1}{2} A + \right. 
\nonumber\\
&+&\left. \frac{\gamma+ \frac{1}{3}}{(-1 + \gamma)(3 A^{1 + 3\gamma} \kappa 
- 8 G M_\gamma  \pi)}
\left(  A G M_\gamma \sqrt{2\pi}\sqrt{ \frac{-3 A^{1 + 3\gamma} \kappa + 8 G 
M_\gamma \pi}{G M_\gamma}} \times \right. \right.\nonumber\\ 
&& \left. \left. \times  
 {_2F_1}\left[ -\frac{3(-1 + \gamma)}{2(1 + 3\gamma)}, \frac{1}{2}, 
 1 - \frac{3(-1 + \gamma)}{2(1 + 3\gamma)},\frac{3 A^{1+ 3\gamma} \kappa}{8 G 
M_\gamma \pi} \right] \right) \right\} ,
\end{eqnarray}
where $_{2}F_{1}$ is the hypergeometric function.

However, we can solve for particular cases of $\gamma$ parameter this integral
(\ref{phi}) as follow
\begin{itemize}
\item{} radiation case:  $\gamma = \frac{1}{3}$  
 \begin{eqnarray}
&&\int \sqrt{384 \pi GM_{1/3}  - 144 \kappa A^2} dA=
\frac{A}{2} \sqrt{ 384 \pi GM_{1/3}- 144 \kappa A^2} \nonumber\\
&+& \frac{16 i G \pi M_{1/3}}{\sqrt{\kappa}} \ln \left\{-24 i A \sqrt{\kappa} +
2 \sqrt{384 \pi GM_{1/3}- 144 \kappa A^2 } \right\},
\end{eqnarray}

\item{} dust fluid: $\gamma =0 $ 
\begin{eqnarray}
&&\int \sqrt{384 \pi GM_0 A - 144 \kappa A^2} dA=
\left(\frac{A}{2} - \frac{2 G \pi M_0}{3\kappa}\right)\sqrt{ 384 \pi GM_0 A- 
144 \kappa A^2} \nonumber\\
&-&\frac{16G^2 M_0^2\pi^2 \sqrt{384 \pi GM_0 A - 144 \kappa A^2 } \ln\left(
2\sqrt{3\kappa A}+2\sqrt{3\kappa A - 8G\pi M_0}\right)}{3\sqrt{3\kappa^3 A}
\sqrt{3\kappa A-8G\pi M_0}},
\end{eqnarray}

\item{} inflation like  case: $\gamma = -1 $ 
\begin{equation}
\int \sqrt{384 \pi GM_{-1} A^4 - 144 \kappa A^2} dA=\frac{1}{2A}
\left(\frac{2A^2}{3} - \frac{\kappa}{4G \pi M_{-1}} \right) 
\sqrt{384 \pi GM_{-1} A^4 - 144 \kappa A^2},
\end{equation}

\item{} stiff fluid: $\gamma = 1 $ 
\begin{eqnarray}
&&\int \sqrt{384 \pi GM_1 A^{-2} - 144 \kappa A^2} dA=
\sqrt{384 \pi GM_1 A^{-2} - 144 \kappa A^2} \left\{ \frac{1}{2A}  + \right. 
\nonumber\\
&+& \left. \frac{i A \sqrt{2\pi G M_1}}{\sqrt{3\kappa A^4-8\pi G M_1}} 
 \ln \left(-\frac{4i \sqrt{ 2\pi G M_1}}{A^2} 
+ \frac{2\sqrt{3\kappa A^4-8\pi G M_1}}{A^2}  \right) \right\}.
\end{eqnarray}
\end{itemize}
These results will be used in the next section, to obtain the solution
according to the supersymmetric quantum mechanics scheme.
\section{Supersymmetric quantum solutions}

To include the factor ordering problem, we substitute the following 
relation into (\ref{hamiltonian1}),
\begin{equation}
A^{-1} \frac{d^2 \Psi}{d A^2} \rightarrow 
A^{-1}\left( \frac{d^2 \Psi}{d A^2} - pA^{-1} \frac{d \Psi}{dA} \right ),
\end{equation}
where the real parameter $p$ measures the ambiguity in the factor ordering.
So, the Wheeler-DeWitt equation, can be written as follows
\begin{equation}
- A\frac{d^2\Psi}{dA^2} + p \frac{d \Psi}{dA} - V(A)\Psi=0,
\label{WDW}
\end{equation}
with $V(A)=384 \pi G M_\gamma A^{-3\gamma + 2} - 144 \kappa A^3$

In this scheme we start giving the following super-hamiltonian
\begin{equation}
H_{super}:= \left( {\cal H}_0+ F 
\frac{\partial^2 \Sigma(A)}{\partial q^\nu \partial q^\mu}\left[
\bar \psi,\psi \right] \right),
\label{supernew}
\end{equation}
where the bosonic hamiltonian ${\cal H}_0$  correspond to the one in
eqn. (\ref{WDW}), F is a complex function and $\Sigma$ is the
superpotential function.
We write the super-charges as follow
\begin{eqnarray}
Q &=&\psi\left( f(A)\frac{d}{dA}+i\frac{d \Sigma(A)}{dA}\right),
\label{carga1}\\
\bar Q &=&\bar \psi\left(f(A) \frac{d}{dA}-i\frac{d \Sigma(A)}{dA}\right).
\label{carga2} 
\end{eqnarray}
where $f(A)$ is an auxiliary function to be determined via the analogy with the
hamiltonian under study.

We suppose the following algebra for the variables \cite{socorro2} 
$\psi$ and $\bar \psi$, 
\begin{equation}
\left\{ \psi ,\bar \psi \right \} =  -1, \qquad 
\left\{ \psi, \psi \right \} = 0, \qquad 
\left\{ \bar \psi,\bar \psi \right \} =0.
\end{equation}
 Using the representation  $\psi=- \frac{d}{d\theta^0}$ 
and $\bar \psi=\theta^0$, one find the
superspace hamiltonian to be written in the form
\begin{eqnarray}
H_{super}\Psi&=&\left\{Q,\bar Q\right\}\Psi=\left(Q\bar Q+\bar QQ\right)\Psi
\nonumber\\
&=&\left(- f^2(A)\frac{d^2}{dA^2} - f(A) \frac{df}{dA} \frac{d}{dA}
- \left( \frac{d\Sigma(A)}{dA}\right)^2 \right. \nonumber\\
&+&\left.  i f(A) \frac{d^2 \Sigma(A)}{dA^2} \left[\bar \psi,\psi \right]  
\right)\Psi .
\label{ss}
\end{eqnarray}
This equation is similar to (\ref{supernew}).

Making the comparation between (\ref{WDW}) and (\ref{ss}) we obtains the 
following relations
\begin{equation}
f^2(A)= A, \qquad p=-\frac{1}{2}, \qquad 
V(A)= \left(\frac{d\Sigma(A)}{dA}\right)^2.
\label{potencial1} 
\end{equation}
We can see that the parameter that give the measure of  the factor ordering is
 fixed at $p=-\frac{1}{2}$ in this approach, leading to
\begin{equation}
- f(A) \frac{df}{dA}\equiv p.
\label{factor} 
\end{equation}
Then, any hamiltonian equation in one dimension that obey this relation,
will have the factor ordering fixed in the supersymmetric regime. In other 
cases, it will be necessary to study the particular hamiltonian equation and 
the supersymmetric scheme.

Moreover, in this scheme, any physical state must obey the following quantum 
constraints

\begin{eqnarray}
\bar Q \Psi=0, \label{qbar} \\
Q\Psi =0.\label{qnobar}
\end{eqnarray}
The wave function has the following decomposition in the Grassmann
variables representation
\begin{equation}
\Psi={\cal u}_+ + {\cal u}_- \theta^0,
\label{funpsi}
\end{equation}
where the component ${\cal u}_+$ is the contribution of the bosonic sector, 
and whereas , ${\cal u}_-$ is the contribution  of the  fermionic sector.

The supercharges read as 

\begin{eqnarray}
Q&=&-\left(\sqrt{A}\frac{d}{dA}+iD_A \Sigma\right)\frac{d}{d \theta^0},\\
\bar Q &=&\theta^0\left(\sqrt{A}\frac{d}{dA}-iD_A \Sigma\right),
\end{eqnarray}
where $D_A= \frac{d}{dA}$.

Using Eq. (\ref{qbar}), we get the following differential equation 
\begin{equation}
 \left( \sqrt{A}\frac{d {\cal u}_+}{dA} -iD_a \Sigma {\cal u}_+ \right) =0. 
\label{uno} 
\end{equation}  

The solution of the latter equation is

\begin{equation}
{\cal u}_+ = {\cal u}_{0+} e^{ i\int \frac{1}{\sqrt{A}}D_A\Sigma \, dA },
\label{w+}
\end{equation}
where ${\cal u}_{0+}$ is an integration constant.

Employing equation (\ref{qnobar}) one gets
\begin{equation}
 \left( \sqrt{A}\frac{d {\cal u}_-}{dA} +iD_a \Sigma {\cal u}_- \right) =0, 
\label{dos} 
\end{equation}  
where ${\cal u}_-$ has the form

\begin{equation}
{\cal u}_- = {\cal u}_{0-} e^{- i\int \frac{1}{\sqrt{A}}D_A\Sigma \, dA }
\label{w2}
\end{equation}

(\ref{w+}) and (\ref{w2})  can be written in the following way
\begin{equation}
{\cal u}_\pm = {\cal u}_{0\pm} e^{\pm i\int \frac{1}{\sqrt{A}}D_A\Sigma \, dA}.
\end{equation}

The integration in these equations corresponds exactly to the
equation (\ref{phi}). Thus, the supersymmetric quantum solutions are obtained 
in closed form.

\section{Conclusions}
In the supersymmetric fashion, the calculation by means of the Grassmann 
variables  of $|\Psi|^2$  given by (\ref{funpsi}) is well known  \cite{Faddeev}
\begin{equation}
(\Psi_1|\Psi_2) = \int{ \left( \Psi_1(\theta^*)\right)^* \Psi_2(\theta^*) 
e^{ -\sum_i \theta_i^* \theta_i^{} }}  \prod_i d\theta_i^* d\theta_i^{},
\label{density}
\end{equation}
where the operation  *  is defined as 
$(C\theta_1^{}...\theta_n^{})^*=\theta_n^*...\theta_1^*C^*$,
with the usual algebra for the Grassmann numbers
$\theta_i \, \theta_j = - \theta_j \, \theta_i $. The rules to integrate over 
these numbers are the following
\begin{equation}
\int{\theta_1^{}\theta_1^*...\theta_n^{}\theta_n^*}d\theta_n^*d\theta_n^{}...
d\theta_1^*d\theta_1^{}=1 
\label{grassman3}
\end{equation}
\begin{equation}
\int d\theta_i^* = \int d\theta_i^{} = 0 .
\label{grassman4}
\end{equation}

In our case, we have $\Psi_1=\Psi_2=\Psi$. So, when we integrate on the
Grassmann numbers, employing also the relations  
 (\ref{grassman3}) and (\ref{grassman4}), we obtain
\begin{equation}
|\Psi|^2 = \bar {\cal u}_+ \, {\cal u}_+^{} + \bar {\cal u}_0 \, {\cal u}_0^{},
\end{equation}
where the $\bar {\cal u}$ symbol means the complex operation.

Using the expresions for the functions  ${\cal u}_+$  and ${\cal u}_-$ given 
in (\ref{w+}) and (\ref{w2}), respectively, we arrive at the following 
expression for the  probability density 
\begin{equation}
|\Psi|^2 =  {\cal u}^2_{0+} + {\cal u}^2_{0-}.
\end{equation}
We thus are able to express (\ref{density}) for our particular problem.
As can be seen, we can infer that the contributions of the bosonic and 
fermionic sectors of the density probability are equal.

The main results in this work are to provide the methodology  to find
the general form for all contributions that ocurring in the expansion  of 
the FRW wave function of the Universe with matter withing the approach of 
Witten's supersymmetric quantum mechanics.
In addition, we find one criterion for fixing the parameter that measure the
factor ordering of the operators. Besides,  we find that the exact solutions 
for the empty (+) and filled (-) sector of the fermion Fock space are at the
same time the lowest-order WKB aproximations (Einstein-Hamilton-Jacobi 
equation). Finally, we find the general form of the  probabilitity density  
(\ref{density}), for the FRW case, including matter fields.  

\noindent {\bf Acknowledgments}\\
We want to thanks H.C. Rosu for critical reading of the manuscript.
This work was partially supported by CONACYT.

\end{document}